\documentclass[amssymb,amsmath]{iopart}
\usepackage{bm}
\usepackage{graphicx}
\usepackage{color}

\begin{document}

\title{A solvable model of Landau quantization breakdown}

\author{T. Champel}
\address{Univ. Grenoble Alpes, CNRS, LPMMC, 25 avenue des Martyrs, 38042 Grenoble, France}

\author{S. Florens}
\address{Univ. Grenoble Alpes, CNRS, Institut N\'{e}el, 25 avenue des Martyrs, 38042 Grenoble, France}



\begin{abstract}
Physics of two-dimensional electron gases under perpendicular magnetic
field often displays three distinct stages when increasing the field amplitude: a low field regime with classical 
magnetotransport, followed  at intermediate field by a Shubnikov-de Haas phase where the transport 
coefficients present quantum oscillations, and, ultimately, the emergence at high field of the quantum Hall effect with 
perfect quantization of the Hall resistance. A rigorous demonstration of
this general paradigm is still limited by the difficulty in solving models of
quantum Hall bars with macroscopic lateral dimensions and smooth disorder. We propose
here the exact solution of a simple model exhibiting similarly two sharp transitions 
that are triggered by the competition of cyclotron motion and potential-induced drift.
As a function of increasing magnetic field, one observes indeed three distinct 
phases showing respectively fully broken, partially smeared, or perfect Landau level 
quantization. This model is based on a non-rotationally invariant, inverted
two-dimensional harmonic potential, from which a full quantum solution is 
obtained using 4D phase space quantization. The developed formalism unifies 
all three possible regimes under a single analytical theory, as well as arbitrary 
quadratic potentials, for all magnetic field values.
\end{abstract}


\section{Introduction}

Magneto-transport in two-dimensional (2D) electronic ga\-ses at low temperatures presents 
ubiquitous features that are observed in vastly different classes of systems, from 
semiconducting heterostructures~\cite{RMPklitzing}, to graphene~\cite{Novo,Zhang,QHEgraphene} 
and other carbon based materials, oxide interfaces~\cite{ReviewMnZnO}, and 
topological systems~\cite{QHEtopo}.
While details in the transport characteristics will strongly depend on the peculiarities 
of a given material (for instance the sign of magnetoconductance variation at low field, 
or the value of Hall conductance quantization plateaus), Landau level formation (or its counterpart at 
decreasing field, Landau level breakdown) appears as a very generic phenomenon. Indeed, 
Landau levels start to be witnessed only from an intermediate magnetic field regime, in 
which mild oscillation of thermodynamic and transport coefficients are observed. 
Only in a second range~\cite{Fogler,Floser} of even higher magnetic field does full quantization of the Hall 
conductance fi\-nally emer\-ge, with the Landau level index becoming a good quantum number.

In order to explain these observations, the electronic motion in a perpendicular magnetic 
field and subject to confining or disordered electrostatic potentials has been thoroughly 
studied by many different theoretical methods 
\cite{Laughlin,Halperin,Streda,Thouless1982,Thouless1984,MacDo,Prange,Buttiker1988,Janssen,Huckestein1995,Dmitriev}.
One clear limitation of current theories is their inability to comprehend all regimes 
of magnetic field (from low to intermediate and high) in a unified way, so that the question 
of the sharpness of the transitions between each regime is not easily established on
general grounds. 

Our goal in this paper is to propose an exactly solvable model of Landau quantization 
breakdown that exhibits clearly two sharp transitions. This simple model is based on
a non-rotationally invariant and inverted 2D parabolic potential, that we exactly
solve in two dimensions for all values of the perpendicular magnetic field. Two
mathematically related but physically distinct quadratic potential models are already well-known from
the literature. The first model~\cite{Fock1928,Darwin1931}, often used to describe 
quantum dots, considers a fully confining 2D parabolic potential, and was solved  
around the same time as the Landau states \cite{Landau1930} for the free motion
problem. This solution led to the Fock-Darwin eigenstates ~\cite{Fock1928,Darwin1931}, showing a discrete energy
spectrum for all magnetic field values. Indeed, the effect of finite magnetic field 
amounts to redefine the quantum states while renormalizing the harmonic spectrum. 
The second model, relevant for quantum point contacts, was proposed and solved
decades later by Fer\-tig and Hal\-pe\-rin~\cite{Fertig1987}, who considered quantum
motion in a quadratic saddle point potential. The mathematical solution is here
more involved due to the use of scattering states in a potential that is unbounded 
from below. Again, this model presents the same feature that the physics is weakly
dependent on magnetic field, with tunneling being mostly renormalized by
cyclotron motion~\cite{Buttiker}. In contrast, we will find that the inverted non-rotationally
invariant quadratic po\-ten\-tial, which is re\-le\-vant to des\-cri\-be anti-dots or Coulomb 
impurities at a local level, displays markedly different electronic states from the low to 
intermediate and finally high magnetic field regimes. 
While this 2D inverted parabolic potential model could be solved by wavefunction 
techniques (adapted to each specific magnetic field range), or using more general path 
integral approaches~\cite{Jain1988,Entelis1992,Kagalovsky1996,Tochishita1996}, we propose here 
an analytic and unified phase space solution, that naturally encompasses all field ranges. 
As a matter of fact, this single solution also accounts for the Fock-Darwin wavefunctions 
and Fertig-Halperin scattering states in the case where the sign of two or one curvatures of 
the potential is inversed respectively.
We note that the case of an inverted one-dimensional parabolic potential was recently 
solved using special functions~\cite{Krason}, and displays a similar albeit simpler phenomenology
compared to the two-dimensional situation with a finite magnetic field.

The approach that we follow here extends previous phase space quantization ideas
\cite{Malkin1969,Champel2007} that were used to derive semiclassical approximations 
in the large magnetic field limit
\cite{Champel2008,Champel2009rapid,Champel2009,Champel2010}, allowing good understanding 
of local density of states measurements \cite{Hashi2008,Hashi2012} in the
quantum Hall regime. This formalism, best suitable at high field, relies upon
wavefunctions that are eigenstates of the free Hamiltonian (with pure Landau level 
spectrum), while maintaining a semi-classical behavior through their coherent state 
character with respect to the guiding center coordinate (displaying hence a 2D
phase space).
This property allows one to easily perform the projection of any states of the
Hilbert space onto a given arbitrary Landau level, a procedure which assumes a 
full energy decoupling between the orbital and guiding center degrees of freedom 
of the electron in the plane. The semiclassical-type approximations are then vindicated 
at high magnetic fields by the slow dynamics of the guiding center.

The main technical development made in the present paper is an extension of the coherent 
state formalism to account efficiently for Landau level mixing. Such a mixing inherently 
induced by a nonuniform electrostatic potential signals that the Landau level index may 
not be a good quantum number anymore, and that the cyclotron motion can in
general not be treated independently of the guiding center motion. 
The key physical insight relies on the use of a 4D phase space,
which allows one to treat both orbital motion and guiding center drifting on an
equal footing. This approach relies on the general dynamics of phase space 
distribution functions, which has been proved to be an autonomous formulation of quantum
mechanics \cite{Bayen1978,Zachos}.
After establishing the general formalism describing the dynamics
for the corresponding 4D Wigner functions, we obtain a unique compact analytical
solution for the case of an arbitrary quadratic potential, accounting for the
three different physical situations discussed above (quantum dots, quantum point 
contacts, quantum antidots). The case of a quantum antidot is found to
display rich physics as a function of magnetic field. The strong magnetic field
regime where both orbital and guiding center motions are associated with
discrete energy levels gets substituted  below a critical magnetic field by a
regime with broadened Landau and antidot energy levels, before an
ultimate breakdown of guiding center and orbital motions at very low
magnetic fields. The behavior shown by our toy model of quantum antidot is
clearly relevant for the understanding of the quantum Hall effect breakdown
occuring in the more complicated case of a random potential.

The plan is organised as follows. 
For completeness, Sec.~2 briefly reviews the high
magnetic field coherent state formalism and the general equations determining 
the electronic motion in the corresponding 2D coherent state representation.
Since the Landau level index is not a good quantum number at any finite
magnetic fields, it becomes relevant in general to replace this
discrete quantum number by an extra continuous degree of freedom with coherent
state character, in the same way as for the guiding center degree of freedom. 
This leads us to work preferentially in a full 4D 
phase space representation provided by a basis of doubly coherent states
\cite{Feldman1970,Varro1984,Manko2012,Zhebrak2013}. Sec.~3 provides the derivation 
of the general equation describing electronic motion in the plane under a
perpendicular magnetic field within a 4D phase space representation
(technical details are provided in Appendix A). The full quantum equation is solved in Sec.~4 for the 
case of arbitrary quadratic electrostatic potentials in terms of two independent 
{\em effective} cyclotron and guiding center motions. Remarkably, in the full
4D phase-space representation, all types of quadratic potential enjoy a generic (unique) 
and compact exact quantum solution, which is valid at any magnetic fields. 
The physics of the seldom considered inverted parabolic potential is investigated 
in Sec.~5, as a model of Landau quantization breakdown.

\section{Review of the 2D-coherent state representation}

In this paper, we consider a single electron of charge $e=-|e|$ and effective
mass $m^{\ast}$ at position ${\bf r}=(x,y)$ in a two-dimensional plane subject
to a perpendicular uniform magnetic field ${\bf B}=B \hat{{\bf z}}$ and an
electrostatic potential $V({\bf r})$. The Hamiltonian reads
\begin{eqnarray}
H=\frac{1}{2 m^{\ast}} \left(-i \hbar {\bm \nabla}_{{\bf r}} -e{\bf A}({\bf r}) \right)^2
+V({\bf r}), \label{Ham}
\end{eqnarray}
where the vector potential ${\bf A}({\bf r})$ is related to the magnetic field
with the equation ${\bm \nabla} \times {\bf A}={\bf B}$. In the absence of
potential [i.e., for $V({\bf r})=0$], the corresponding quantum mechanical
problem can be readily solved and yields the well-known quantization of the
kinetic energy into discrete Landau levels $E_{n}=(n+1/2) \hbar \omega_c$ where
$n$ is a positive integer (a.k.a. the Landau level index) and $\omega_c$ is the
cyclotron frequency proportional to the magnetic field amplitude $B$ as
$\omega_c=|e|B/m ^{\ast}$. Owing to the degeneracy of the kinetic energy levels,
it is possible to showcase different bases of eigenstates associated with this
Landau level quantization. 

A physically transparent basis is provided by the vortex set of states
\cite{Malkin1969,Champel2007} expressed in the symmetrical gauge ${\bf A}={\bf
B} \times {\bf r}/2$ as 
\begin{eqnarray}
\langle{\bf r} | n,{\bf R} \rangle = \frac{1}{l_B \sqrt{2 \pi n!}} 
\left(\frac{z-Z}{\sqrt{2} l_B} \right)^{n} \, e^{-\frac{|z|^2+|Z|^2-2 Z z^{\ast}}{4 l_B^2}} 
, \label{vortex}
\end{eqnarray}
where $l_B=\sqrt{\hbar/|e|B}$ is the magnetic length and $z=x+iy$ refers to the
electron position in the complex plane. Within this peculiar set of eigenstates
of the Landau level problem, the degeneracy quantum number is provided by the
vortex position ${\bf R}=(X,Y)$, associated with the complex coordinate $Z=X+iY$
in the complex plane, which uniquely characterizes for $n \geq 1$ the location of the
zeros of the wave function in the two-dimensional plane. In the limit of
vanishing $l_B$ the positions ${\bf R}$ reduce to the classical guiding center
location. Despite presenting a nonorthogonal overlap with respect to the quantum
number ${\bf R}$ typical of coherent states
\begin{eqnarray}
\langle n_1,{\bf R}_1| n_2,{\bf R}_2 \rangle= \delta_{n_1,n_2}\, e^{-\frac{|Z_1|^2+|Z_2|^2-2Z_1^{\ast} Z_2}{4 l_B^2}} 
,
\end{eqnarray}
the states (\ref{vortex}) form a coherent state basis with respect to the
guiding center coordinate (within each Landau level), obeying the completeness relation
\begin{eqnarray}
\int \!\! \frac{d^2 {\bf R}}{2\pi l_B^2} \sum_{n=0}^{+\infty} |n,{\bf R} \rangle \langle n,{\bf R} |=1.
\end{eqnarray}
By associating the incremental area $d^2{\bf R}$ with the area $2\pi l_B^2$,
this relation explicitly points out the degeneracy of the Landau levels to be
$(2 \pi l_B^2)^{-1}$ per unit area.

This Landau level degeneracy gets lifted when considering a non-uniform
potential $V({\bf r})$. At high magnetic fields, i.e., when Landau level mixing
can reasonably be neglected,
 the
degeneracy lifting process becomes nonperturbative in nature and is the source
of theoretical difficulties. The continuous character of the degeneracy quantum
number ${\bf R}$ in the vortex state basis $|n,{\bf R}\rangle$ then offers a
differential perspective of this process by an arbitrary potential, which has
been thoroughly studied during the last decade in a series of papers
\cite{Champel2008,Champel2009,Champel2010}. Due to the coherent state nature of
the degree of freedom ${\bf R}$, the electronic Green's function in the time
domain $t$ corresponding to Hamiltonian (\ref{Ham}) can be written as the
convolution
\begin{eqnarray}
G({\bf r},{\bf r'};t)=\int \!\! \frac{d^2 {\bf R}}{2 \pi l_B^2} \sum_{n_1,n_2} K_{n_1,n_2}({\bf r},{\bf r}';{\bf R}) \, g_{n_1,n_2}({\bf R};t) \label{Green}
\end{eqnarray}
where the electronic structure factor defined by
\begin{eqnarray}
K_{n_1,n_2}({\bf r},{\bf r}'; {\bf R})=e^{-(l_B^2/4) \Delta_{{\bf R}}} \left[ \langle n_2,{\bf R}|{\bf r}' \rangle \, \langle {\bf r} | n_1,{\bf R}\rangle \right]
\end{eqnarray}
is independent of the electrostatic potential $V({\bf r})$ and
embodies the quantum contribution arising from the pure orbital motion of the
electron (here $\Delta_{{\bf R}}$ is the Laplacian operator taken with respect
to the position ${\bf R}$). The vortex Green's function components
$g_{n_1,n_2}({\bf R};t)$, which encode the quantum drift of the guiding center
induced by $V({\bf r})$, obey the equations
\begin{eqnarray}
\hspace*{-2cm}
\left(i\hbar \partial_{t}-E_{n_1} \pm i 0^+\right) g_{n_1,n_2}({\bf R};t) 
- \sum_{n_3} v_{n_1,n_3}({\bf R}) \star_{{\bf R}} g_{n_3,n_2}({\bf R};t)=\delta_{n_1,n_2} \delta(t)
\label{Dyson}
\end{eqnarray}
with the effective potential matrix elements 
\begin{eqnarray}
v_{n_1,n_2}({\bf R})=\int \!\! d^2 {\bf r} \, K_{n_1,n_2}({\bf r},{\bf r};{\bf R}) \, V\left({\bf r} \right) \label{pot}
\end{eqnarray}
expressing the average of the bare potential $V({\bf r})$ over the quantized
orbital motion. Here the infinitesimal quantity $\pm 0^{+}$ relates to the
retarded or advanced Green's functions. The symbol $\star_{{\bf R}}$ is a
pseudodifferential infinite-order symplectic operator
\begin{eqnarray}
\star_{{\bf R}}=\exp\left[i \frac{l_B^2}{2}\left(\overleftarrow{\partial}_{X}\overrightarrow{\partial}_{Y} -\overleftarrow{\partial}_{Y}\overrightarrow{\partial}_{X} \right) \right], \label{star}
\end{eqnarray}
where the arrows above the partial derivatives indicate to which side (left or
right) they have to be applied. It is a magnetic version of the Groenewold-Moyal
star product \cite{Zachos}, with $l_B^2$ playing the role of an effective
Planck's constant and the one-dimensional conjugated variables, position and
momentum, being replaced by the components $X$ and $Y$ of the orbit center in
the two-dimensional plane.

The exact expression (\ref{Green}) translates into the quantum mechanical
language the natural decomposition of the electronic motion 
into orbital and orbit center degrees of freedom. The vortex representation
introducing both discrete and continuous quantum numbers turns out to be
well-suited to treat quantitatively the resulting electronic dynamics at high
magnetic fields, since it structurally encodes that these two elementary
motions are characterized by very different time scales: the fast orbital degree
of freedom is described in discrete terms, while a continuous classical phase
space representation of the Landau level degeneracy is vindicated by the slow
dynamics of the orbit center. In the high magnetic field regime, a good
(perturbative) approximation is to entirely separate these two time scales by
considering that the orbital motion gets decoupled from the guiding center
motion. Technically, this implies restricting the electron dynamics to a given
Landau level subspace. This state projection is conveniently performed for any
Landau levels through the analyticity property of the vortex state basis
(\ref{vortex}) in the complex guiding center variable $Z$, which holds
irrespective of the Landau level index $n$ (in contrast, the well-known
anti-analyticity property of the wave functions in the electronic variable $z$
only holds for the lowest Landau level). In terms of vortex Green's functions,
only diagonal elements $g_{n_1,n_1}({\bf R};t)$ contribute to the overall
electron dynamics in expression (\ref{Green}) after projection. Then, at the
level of the guiding center motion, the star product operator (\ref{star})
generates a hierarchy of local energy scales ordered by powers of $l_B^2$ and
successive spatial derivatives of the effective potential (\ref{pot}), which
allows one to devise semiclassical nonperturbative approximation schemes for the
vortex Green's functions $g_{n_1,n_1}({\bf R};t)$ valid at small times $t$ (and
physically justified at finite temperatures).
 
The objective of this paper is to address the situation beyond the Landau level
projection, i.e., to eventually relax the high magnetic field constraint. This
means to deal in Eq. (\ref{Dyson}) with the entire matrix structure of the
vortex Green's functions associated to the Landau levels together with the
differential aspects related to the guiding center dependence. 
In the following, we shall develop an alternative strategy valid at any magnetic
fields, which requires a reformulation of the quantum representation of the
states.


\section{General equation of motion in the 4D-coherent state representation}
The idea is to treat the two electronic degrees of freedom associated to the
cyclotron motion and the guiding center motion on an equal footing, i.e., within
a fully differential 4-dimensional phase-space perspective. 
For this purpose, we introduce a coherent state representation of the orbital
degree of freedom by defining the doubly coherent states $|{\bm \rho},{\bf R}
\rangle$, built from the vortex states as
\begin{eqnarray}
|{\bm \rho},{\bf R} \rangle = e^{- \frac{|\zeta|^2}{4 l_B^2}} \, \sum_{n=0}^{+ \infty} \frac{1}{\sqrt{n!}} \left(\frac{\zeta}{\sqrt{2}l_B }\right)^n | n,{\bf R} \rangle, \label{defrho}
\end{eqnarray}
where the orbital position ${\bm \rho}=(\rho_x,\rho_y)$ replaces the quantized Landau level 
index $n$ by a continuous cyclotron motion around the guiding center ${\bm R}$ in the two-dimensional 
plane, with $\zeta=\rho_x+i \rho_y$ its complex number representation (thus
$\zeta^{\ast}=\rho_x-i \rho_y$). It can be easily established that this set of
states form a bi-coherent states basis, with the standard non-orthogonal 
overlap expression:
\begin{eqnarray}
 \hspace*{-1.5cm} \langle {\bm \rho}_1 ,{\bf R}_1| {\bm \rho}_2 ,{\bf R}_2 \rangle 
=  \langle {\bm \rho}_1 | {\bm \rho}_2 \rangle \, \langle {\bf R}_1| {\bf R}_2 \rangle 
=
e^{-\frac{|\zeta_1|^2+|\zeta_2|^2-2\zeta_1^{\ast} \zeta_2 }{4 l_B^2} } \, 
e^{-\frac{|Z_1|^2+|Z_2|^2-2Z_1^{\ast} Z_2 }{4 l_B^2} } 
, \nonumber \\
\end{eqnarray}
and the completeness relation 
\begin{eqnarray}
 \int \!\! \frac{d^2 {\bf R}}{2 \pi l_B^2} \int \!\! \frac{d^2 {\bm \rho}}{2\pi l_B^2} | {\bm \rho} 
,{\bf R} \rangle \langle {\bm \rho},{\bf R} |=1 .
\end{eqnarray}
From Eqs. (\ref{vortex}) and (\ref{defrho}) one easily gets the expression for
the fully coherent wave function (which already appeared in the literature several decades ago, see e.g. Refs. \cite{Feldman1970,Varro1984})
\begin{eqnarray}
\langle {\bf r} | {\bm \rho},{\bf R} \rangle= \frac{1}{l_B \sqrt{2 \pi}} \, 
e^{-\frac{|z|^2+|Z|^2+|\zeta|^2-2Z z^{\ast}-2 \zeta (z-Z)}{4 l_B ^2}}.
\end{eqnarray}

The corresponding Green's functions in this representation of bi-coherent states
are obtained from the vortex Green's functions components via a simple change of
basis as
\begin{eqnarray}
\hspace*{-1.5cm} 
 g_{{\bm \rho}_1,{\bm \rho}_2}\left({\bf R};t\right)= e^{- \frac{|\zeta_1|^2+|\zeta_2|^2}{4 l_B^2}} 
\,\sum_{n_1=0}^{+\infty} \sum_{n_2=0}^{+\infty} \left(\frac{\zeta_1^{\ast}}{\sqrt{2}l_B }\right)^{n_1} 
\left(\frac{\zeta_2}{\sqrt{2}l_B }\right)^{n_2} \frac{g_{n_1,n_2}({\bf R};t)}{\sqrt{n_1!\, n_2!}}. 
\label{defgrho}
\end{eqnarray}
The analytical dependence of these functions on the variables $\zeta_1^{\ast}$ and
$\zeta_2$ is put to good use in order to write down a general
``diagonal'' expression for the electronic Green's function (see Appendix A for
a detailed derivation) similarly to Eq. (\ref{Green})
\begin{eqnarray}
G({\bf r},{\bf r'};t)=\int \!\! \frac{d^2 {\bf R}}{2 \pi l_B^2} \int \!\! \frac{d^2 {\bm \rho}}{2 \pi l_B^2} K({\bf r},{\bf r}';{\bm \rho},{\bf R}) \, g({\bm \rho},{\bf R};t) \label{Green2}  \hspace*{0.5cm}
\end{eqnarray}
with the Kernel function 
\begin{eqnarray}
K({\bf r},{\bf r}';{\bm \rho},{\bf R})=e^{-(l_B^2/4) \left(\Delta_{{\bf R}}+\Delta_{{\bm \rho}} \right)} \left[ \langle {\bm \rho},{\bf R}|{\bf r}' \rangle \, \langle {\bf r} | {\bm \rho},{\bf R}\rangle \right],  \hspace*{0.5cm}
\end{eqnarray}
and where the diagonal component functions $g({\bm \rho},{\bf R};t)$ obey the relatively compact (exact) equation 
\begin{eqnarray}
 \left(i \hbar \partial_{t} \pm i 0^+\right) g({\bm \rho},{\bf R};t) - E({\bm \rho},{\bf R}) 
\star_{{\bf R}} \star_{{\bm \rho}} \, g({\bm \rho},{\bf R};t)=\delta(t). \nonumber \\
\label{Dysonrho}
\end{eqnarray} 
The matrix structure encountered into the previous system of equations (\ref{Dyson}) has been replaced in the present four-dimensional phase space representation by the presence of an additional pseudodifferential infinite-order symplectic operator
\begin{eqnarray}
\star_{{\bm \rho}}=\exp\left[i \frac{l_B^2}{2}\left(\overleftarrow{\partial}_{\!\! \rho_x}\overrightarrow{\partial}_{\!\! \rho_y} -\overleftarrow{\partial}_{\!\! \rho_y}\overrightarrow{\partial}_{\!\! \rho_x} \right) \right], \label{starrho}
\end{eqnarray}
whose structure is the same as that of the star product operator $\star_{\bf{R}}$ which governs the quantum motion of the guiding center ${\bf R}$.
Here, the quantity $E({\bm \rho},{\bf R})$ expresses the classical total energy 
\begin{eqnarray}
E({\bm \rho},{\bf R})=\frac{1}{2} m^{\ast}\omega_c^2 {\bm \rho}^2 + V\left({\bm \rho}+{\bf R} \right)
, \label{potrho}
\end{eqnarray}
which includes both the classical (rotational) kinetic energy contribution
(which was previously associated with the Landau levels) and the potential
energy contribution. Expression~(\ref{potrho}) is obvious on semi-classical
grounds, and the only difficulty brought by quantum mechanics in its 4D phase
space representation is the necessity to deal with the star-product~(\ref{starrho}).

The full phase space formulation provided by the use of the bi-coherent state
set thus offers a physically transparent perspective, with the explicit
implementation of the electron motion decomposition ${\bf r}={\bm \rho}+{\bf R}$
in the quantum realm. Note that the 4D phase space is characterized here by
two spatial coordinates ${\bm \rho}$ and ${\bm R}$, in contrast to the more standard
phase space representation with electronic coordinate ${\bm r}$ and its zero-field 
conjugate momentum ${\bm p = -i\hbar \nabla}$. In fact, under a finite magnetic
field, the canonical quantization readily shows that $(X,Y)$ and
$(\rho_x,\rho_y)$ each constitute a quantum conjugate pair, vindicating our
choice of 4D phase space representation. As a consequence,
the main difficulty in this
deformation quantization formulation is entirely embodied in the infinite-order
differential operators $\star_{{\bf R}}$ and $\star_{{\bm \rho}}$ appearing in
Eq. (\ref{Dysonrho}). In general, the electronic potential energy term $V({\bf
r})$ introduces a coupling between the orbital ${\bm \rho}$ and the guiding
center ${\bf R}$ degrees of freedom, which makes this quantum problem
generically quite complicated to solve. Nevertheless, as shown in the next
section, an exact decoupling can be handled for any quadratic potentials.

\section{Generic solution for arbitrary quadratic potentials}

So far, we have derived the general quantum equation (\ref{Dysonrho}) obeyed by the Green's
functions in the 4D phase space representation, without resorting to any specific
form for the potential $V({\bf r})$. The case of a linear potential term does
not present peculiar difficulties, since it does not lead to a coupling between
the orbital and guiding center degrees of freedom. Consequently, from now on we
focus on the case of quadratic potentials which can be written without loss of
generality (a translation and a rotation of the coordinates lead immediately
to the most generic quadratic form) as
\begin{eqnarray}
V({\bf r})=a x^2 + b y^2, \label{potV}
\end{eqnarray}
where $a$ and $b$ are arbitrary real coefficients, which encompass the three     
possible cases of potentials: i) confining (parabolic case, $a>0$ and $b>0$);
ii) saddle point (hyperbolic case, $ab<0$); iii) impurity-like (inverted parabolic case,
$a<0$ and $b<0$).
Therefore, the total energy (\ref{potrho}) reads
\begin{eqnarray}
E({\bm \rho},{\bf R})=a (X+\rho_x)^2+b(Y+\rho_y)^2+c {\bm \rho}^2
\end{eqnarray}
with $c=\frac{1}{2} m^{\ast} \omega_c^2$. The difficulty obviously comes from
the presence of terms mixing the ${\bm \rho}$ and ${\bf R}$ coordinates, a
hallmark of quadratic (squared) contributions.

The above equation (\ref{Dysonrho}) for the phase space Green's functions can be solved exactly through the
introduction of a well-chosen change in variables $({\bm \rho},{\bf R}) \to
({\bf R}_1,{\bf R}_2)$, which allows us to simultaneously decouple the spatial
dependences in the total energy and in the differential star-operators. More
explicitly, we impose that the total energy reads after the variable
transformation as
\begin{eqnarray}
E({\bm \rho},{\bf R})=V_1({\bf R}_1)+V_2({\bf R}_2),
\label{cons}
\end{eqnarray}
where the new (quadratic) potential functions $V_1$ and $V_2$ will be determined
later on. A second condition is that the new star-products $\star_{{\bf R}_1}$
and $\star_{{\bf R}_2}$ defined with respect to the new variables ${\bf R}_1$
and ${\bf R}_2$ remain decoupled (typically, we do not want to generate
cross-derivative terms like $\overleftarrow{\partial}_{\!\! X_1}
\overrightarrow{\partial}_{\!\! Y_2}$). The solutions to the differential
equations in ${\bf R}_1$ and ${\bf R}_2$ are then derived separately (without
the $\delta(t)$ source term) and are appointed in the following as the functions
$f_1({\bf R}_1;t)$ and $f_2({\bf R}_2;t)$ with the property that $f_1({\bf
R}_1;0)=f_2({\bf R}_2;0)=1$. Hence, it can be easily shown that the full
solution of Eq. (\ref{Dysonrho}) is given by the product function 
\begin{eqnarray}
\label{g4D}
g({\bm \rho},{\bf R};t)= \mp i \theta(\pm t) \,f_1({\bf R}_1;t) f_2({\bf R}_2;t) \label{fullsolution}
\end{eqnarray}
with $\theta(t)$ the Heaviside step function, and where the functions
$f_{j}({\bf R}_j;t)$ with $j=1$ or $2$ obey the equation
\begin{eqnarray}
 \left(i \hbar  \partial_{t} \pm i 0^+\right) f_j({\bf R}_j;t) - V_j({\bf R}_j) \star_{{\bf R}_j} \, f_j({\bf R}_j;t)=0.  \hspace*{0.5cm}
\label{equationf}
\end{eqnarray} 
This latter equation is very similar to the one obtained for the pure
(decoupled) guiding center motion at high magnetic fields after Landau level
projection. We can thus follow the derivation detailed in Ref.
\cite{Champel2009} to directly write down the solution
\begin{eqnarray}
\label{fRt}
f_j({\bf R}_j;t)=\frac{e^{-i[V_j({\bf R}_j)-V_j({\bf R}_{j0})] \tau_j(t)}}
{\cos \left[\sqrt{\gamma_j} t/\hbar \right]} \, e^{-\frac{it}{\hbar} \left[ V_j({\bf R}_{j0}) \mp i0^+\right]} \label{solution},  \hspace*{0.5cm}
\end{eqnarray}
where 
\begin{eqnarray}
\tau_j(t)=\frac{1}{\sqrt{\gamma_j}} \tan \left( \sqrt{\gamma_j} t /\hbar \right).
\label{tau}
\end{eqnarray}
Here the point ${\bf R}_{j0}$ refers  to the critical point of the qua\-dra\-tic
potential $V_j$, i.e., $\left. {\bm \nabla}_{{\bf R}_j}V_j({\bf R}_j)
\right|_{{\bf R}_{j}={\bf R}_{j0}}={\bf 0}$, and the (uniform) quantity
$\gamma_j$ is related to the Gaussian curvature of the potential $V_j$ as
\begin{eqnarray}
\gamma_j= \frac{l_B^4}{4} \left[\partial^2_{X_j}V_j \partial^2_{Y_j}V_j -\left(\partial_{X_j} \partial_{Y_j} V_j   \right)^2 \right].
\end{eqnarray}
This quantity plays a pivotal role, since its square root crucially determines the relative
time dependence of the Green's function, and thus the spectral properties of the
electronic motion. For instance, when $\gamma_j$ is real positive, the function
$ f_j({\bf R}_j;t)$ contains a periodical dependence in time, which can be
restated as a Fourier series expansion to yield the alternative expression
\begin{eqnarray}
 f_j({\bf R}_j;t)=\sum_{p_j=-\infty}^{+ \infty} a_{p_j}({\bf R}_j)  \, e^{\frac{it}{\hbar} \left[p_j \sqrt{\gamma_j}- V_j({\bf R}_{j0}) \pm i0^+\right]}. \label{fconfin}
\end{eqnarray}
It has been shown in Appendix A of Ref. \cite{Champel2010} that the series coefficients  read
\begin{eqnarray}
a_{p_j}({\bf R}_j) &=&2 (-1)^{n_j} e^{-|\rho_j ({\bf R}_j) |} L_n \left(2|\rho_j ({\bf R}_j) | \right), \\
 \rho_j ({\bf R}_j) &=&\frac{V_j({\bf R}_j)-V_j({\bf R}_{j0})}{\sqrt{\gamma_j}},
\end{eqnarray}
whenever $p_j=\chi_j (2n_j+1)$ with $n_j$ a positive integer and $\chi_j=\pm 1$
whether the potential $V_j$ is convex or concave (here  $L_n(x)$ is the Laguerre
polynomial of degree $n$), and $a_{p_j}({\bf R}_j)=0$ whenever $p_j \neq \chi_j
(2n_j+1)$. From expression (\ref{fconfin}) valid when $\gamma_j \geq 0$ it is
thus readily understood that the energy contribution arising from the potential
$V_j$ is quantized with energy gaps given by $2 \sqrt{\gamma_j}$. For
$\gamma_j<0$, it is understood  in Eqs. (\ref{solution}) and (\ref{tau})  that
$\sqrt{\gamma_j}=i \sqrt{-\gamma_j}$, so that the cosine and tangent
trigonometric functions transform into their hyperbolic counterparts. As a
result, the time periodicity of the Green's function is replaced by a decay on
the time scale $1/\sqrt{-\gamma_j}$ due to the cutoff function
$1/\cosh(\sqrt{-\gamma_j}t)$, which can be seen as a manifestation of quantum
tunneling effects.

From the above requirements on the variables decoupling, it is clear that a linear transformation of the coordinates will fit our purpose. Let us write the original variables in terms of the new ones as
\begin{eqnarray}
X &=& \lambda(X_1+\alpha X_2) , \hspace*{0.5cm}
\rho_x = \lambda(X_2 + \beta X_1), \nonumber \\
Y &=& \lambda(Y_1+ \eta Y_2), \hspace*{0.75cm}
\rho_y = \lambda(Y_2 + \delta Y_1),
\end{eqnarray}
with $\lambda>0$.
The condition for the absence of cross-terms in the star products yields
$\beta=\eta$ and $\alpha=\delta$. Furthermore, we get $
\overleftarrow{\partial}_{\!\! X} \overrightarrow{\partial}_{\!\!
Y}-\overleftarrow{\partial}_{\!\! \rho_x} \overrightarrow{\partial}_{\!\!
\rho_y} = \overleftarrow{\partial}_{\!\! X_1} \overrightarrow{\partial}_{\!\!
Y_1}-\overleftarrow{\partial}_{\!\! X_2} \overrightarrow{\partial}_{\!\! Y_2} $
provided that $\lambda^{-2} =1-\alpha \beta$. The other constraint (\ref{cons})
leads to $\beta= \alpha a/b$ with
\begin{eqnarray}
\alpha=-\frac{1}{2a} \left[a+b+c - \sqrt{(a+b+c)^2-4ab} \right].
\end{eqnarray}
Note that only this combination is compatible with the equality $a \alpha =b
\beta$ (this comes out by considering, e.g., the limit $b \to 0$ which
necessarily implies $\alpha \to 0$). From this, we obtain
\begin{eqnarray}
\lambda^2=\frac{1}{2} \frac{a+b+c + \sqrt{(a+b+c)^2-4ab}}{\sqrt{(a+b+c)^2-4ab}}.
\end{eqnarray}
The effective quadratic potentials read $V_j({\bf R}_j)=a_j X_j^2+b_j Y_j^2$ with
\begin{eqnarray}
\label{a1}
a_1 &=&
\lambda^2 \left[a(1+\beta)^2+c \beta^2 \right] \nonumber \\
&= &
\frac{a}{2b} \left[b-a-c+ \sqrt{(a+b+c)^2-4ab} \right]
, \hspace*{1cm}
\\
\label{b1}
b_1 &=& \lambda^2 \left[b(1+\alpha)^2+c \alpha^2 \right] \nonumber \\
&=&
\frac{b}{2a} \left[a-b-c+ \sqrt{(a+b+c)^2-4ab} \right]
,
\\
\label{a2}
a_2 &=&
\lambda^2
\left[a(1+\alpha)^2+c  \right] \nonumber \\
&=& \frac{1}{2} \left[c+a-b+ \sqrt{(a+b+c)^2-4ab} \right]
, 
\\
\label{b2}
b_2 &=& \lambda^2 \left[b(1+\beta)^2+c  \right] \nonumber \\
&=&
 \frac{1}{2} \left[c+b-a+ \sqrt{(a+b+c)^2-4ab} \right]
.
\end{eqnarray}
The new variables are expressed in terms of the original guiding center and orbital coordinates as
\begin{eqnarray}
\label{X1}
X_1 &=& \lambda(X- \alpha \rho_x), \hspace*{0,5cm}
\label{X2}
X_2 = \lambda(\rho_x- \beta X),  \\
\label{Y1}
Y_1 &=& \lambda(Y-\beta \rho_y),  \hspace*{0,6cm}
\label{Y2}
Y_2 = \lambda(\rho_y- \alpha Y).
\end{eqnarray}
By considering the high magnetic field limit $c \gg |a|,|b|$ for which
$\lambda=1$ and $\alpha=\beta=0$, it is clear that, in general, the degree of
freedom ${\bf R}_1$ plays the role of an effective guiding center, while 
${\bf R}_2$ corresponds to an effective orbital degree of freedom.
The final explicit solution for arbitrary quadratic potentials can be
read off from expression~(\ref{g4D}) for the 4D
phase-space Green's function $g({\bm \rho},{\bm R};t)$, expressed from the
functions $f_j({\bm R_j};t)$ in Eq.~(\ref{fRt}), with the coordinates ${\bm R_j}$
given in Eqs.~(\ref{X1})-(\ref{Y2}) and the effective potentials $V_j({\bm R_j})=a_jX_j^2+b_jY_j^2$ 
determined by the four coefficients in Eqs.~(\ref{a1})-(\ref{b2}). One remarkable
aspect of this general solution is that it does not require the computation of 
any special functions.

\section{Model of Landau quantization breakdown}
Let us finally analyze some physical features of the exact quantum solution
(\ref{fullsolution}). As underlined above, 
the Gaussian curvatures $\gamma_1$ and $\gamma_2$ of the effective 
potentials $V_1$ and $V_2$ are key quantities determining the nature itself 
of the energy spectrum:
\begin{eqnarray}
\label{gamma1}
\gamma_1 &=& l_B^4 a_1 b_1
= l_B^4 \frac{c}{2} \left[a+b+c- \sqrt{(a+b+c)^2-4ab} \right] , \nonumber \\ &&
\\ \label{gamma2}
\gamma_2 &=& l_B^4 a_2 b_2 = l_B^4 \frac{c}{2} \left[a+b+c+ \sqrt{(a+b+c)^2-4ab} \right] . \nonumber \\
\end{eqnarray}
Note that here $l_B^2 c=\frac{1}{2}\hbar \omega_c \geq 0$, while the
coefficients $a$ and $b$ can be chosen positive or negative real numbers depending 
on the spatial configuration for the original potential energy $V({\bf r})$. 

A rich variety of regimes, showing structural changes in the energy spectrum, only takes 
place when both $a$ and $b$ are negative,
corresponding to a toy-model of quantum antidot.
Indeed, in the other circumstances (i.e., for $ab <0$, or for $a$ and $b$ both
positive), one always gets a real positive $\gamma_2$, which signals the
discrete quantization of the effective orbital motion with gaps given by the
energy scale $ \hbar \Omega_c=2 \sqrt{\gamma_2}$. Landau quantization is
thus robust at arbitrary small magnetic field for confining or saddle-point
potentials. The renormalization of the cyclotron frequency from $\omega_c$ to
$\Omega_c$ due to the Landau level mixing processes can be translated into
a renormalization of the magnetic length $l_B$ by introducing the new length
$L=l_B \left(l_B^4 c^2/\gamma_2\right)^{1/4} \equiv l_B \left(\omega_c/\Omega_c 
\right)^{1/2}. $
Moreover, it is instructive to rewrite the curvature of the effective potential
$V_1$ as $\gamma_1=l_B^4 ab l_B^4 c^2/\gamma_2 \equiv \gamma
\left( \omega_c/\Omega_c\right)^2$, 
with $\gamma=l_B^4 ab$ the Gaussian curvature of the electronic
potential energy $V({\bf r})$.
This proportionality relation shows that when $\gamma_2 > 0$ the sign of
$\gamma_1$ dictating the effective guiding center motion is in fact entirely
determined by the bare potential curvature. Nevertheless, the effective guiding 
center follows equipotential lines of the effective potential $V_1({\bf R}_1)$, which,
in the presence of Landau level mixing, differ from those of $V({\bf R}_1)$ and
evolve in magnetic field.

\begin{figure}[ht]
\begin{center}
\resizebox{0.75\textwidth}{!}{ \includegraphics{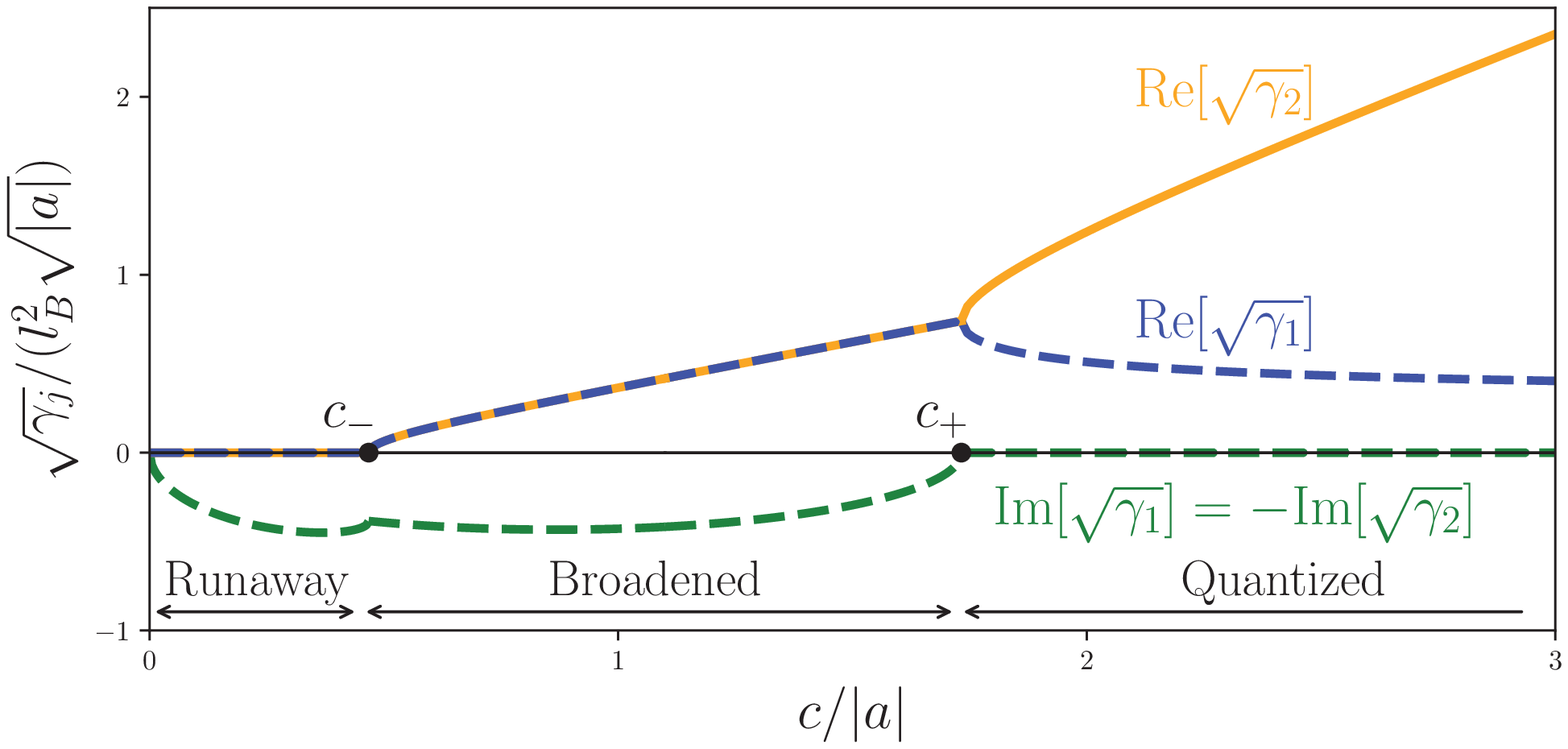}}
\end{center}
\caption{{\bf Dynamical regimes for a non rotationally invariant antidot.}
The figure displays the real and imaginary parts of the quantities $\sqrt{\gamma_1}$ and $\sqrt{\gamma_2}$  characterizing the energy spectrum and related
to the two effective potentials $V_1$ and $V_2$ (defined in Eqs.~(\ref{a1})-(\ref{b2})), which are associated, respectively, to the effective guiding center 
coordinate ${\bm R_1}$ and to the effective orbital motion ${\bm R_2}$, as a
function of the classical cyclotron energy $c=m^{\ast} \omega_c^2/2$.
Here an asymmetric inverted parabolic potential with $a=10 \, b<0$ in Eq. (\ref{potV}) has been considered.
Landau quantization at large magnetic field ($c \geq c_{+}$) corresponds to real and positive potential curvatures,
while the intermediate magnetic field range ($c_{-} \leq c \leq c_{+}$) shows broadened Landau and antidot
levels due to the non-zero imaginary parts of  $\sqrt{\gamma_1}$ and $\sqrt{\gamma_2}$. However, the relation  $\mathrm{Im}[\sqrt{\gamma_1}]=
-\mathrm{Im}[\sqrt{\gamma_2}]$ translates the fact that the effective cyclotron orbits shrink while
the effective guiding center follows orbits that increase as the electron spins down
the potential landscape.
Finally, both curvatures $\gamma_1$ and $\gamma_2$ become negative at low magnetic fields ($c \leq c_{-}$) so that $\sqrt{\gamma_1}$ and $\sqrt{\gamma_2}$ are purely imaginary, corresponding
to the electron running down the inverted parabolic potential without performing any cyclotron motion.}
\label{Curvatures}

\end{figure}

Focusing now the analysis on the inverted parabolic case ($a$ and $b$  negative), it is clear that
both effective potential curvatures $\gamma_1$ and $\gamma_2$ are real and positive for strong enough
magnetic field, as seen by taking the limit of large $c$ in
Eqs.~(\ref{gamma1})-(\ref{gamma2}). Having $\gamma_2>0$ signals robust high magnetic field Landau
quantization, while $\gamma_1>0$ demonstrates that the effective antidot potential $V_1$ confines 
the electronic motion due to the strong Lorentz force, despite the bare antidot potential $V({\bf r})$ of Eq. (\ref{potV}) being repulsive and
unbounded from below. Decreasing the magnetic field, i.e. reducing the value of $c$, one encounters a first critical
value $c_+=(\sqrt{|a|}+\sqrt{|b|})^2$ below which the term under the square root in Eqs.~(\ref{gamma1})-(\ref{gamma2})
becomes negative. In this case, both orbital and guiding center effective motions lock into
decaying orbits (in a semiclassical viewpoint), leading to a finite broadening
of both the Landau and antidot energy levels associated to the finite imaginary parts of the quantities $\sqrt{\gamma_1}$ and $\sqrt{\gamma_2}$. Landau quantization only survives on short
time-scales in this field regime, where cyclotron orbits shrink
($\mathrm{Im}[\sqrt{\gamma_2}]>0$), while the guiding center makes larger and larger
loops around the antidot potential ($\mathrm{Im}[\sqrt{\gamma_1}]=-\mathrm{Im}[\sqrt{\gamma_2}]<0$).
Finally, when $a\neq b$ so that angular momentum is no more conserved, one finds  for lower magnetic fields a
second critical value $c_-=(\sqrt{|a|}-\sqrt{|b|})^2$ below which  both curvatures $\gamma_1$
and $\gamma_2$ become  real and negative, and thus the quantities $\sqrt{\gamma_1}$ and  $\sqrt{\gamma_2}$ get purely imaginary. This low field regime for asymmetric antidot potentials corresponds to the rapid
runaway of the electron down the inverted potential without any looping motion
from the Lorentz force, so that the magnetic orbital effects are totally washed out.
This rich scenario of Landau quantization breakdown in an inverted quadratic potential 
is illustrated on Fig.~\ref{Curvatures} in the case $a=10\, b<0$. 
Note that the Gaussian curvatures in Eqs.
(\ref{gamma1})-(\ref{gamma2}) are purely classical concepts, as the same quantities 
naturally appear when solving the characteristic Newtonian equation of motion of a 
charged particule in an arbitrary quadratic potential in presence of a magnetic field. 
In this sense, the Landau levels breakdown is not intrinsically quantum in nature, although it will 
affect electronic motion at the quantum level.

\section{Conclusion and final remarks}

We have developed a 4D phase space representation of the in-plane electronic
quantum motion in a perpendicular magnetic field, which is relevant beyond the
Landau level projection. While a 2D-coherent state representation considering
the discrete Landau level index as a good quantum number appears still efficient
at moderately small Landau level mixing, the recourse to a bi-coherent state
representation for which both the guiding center and the orbital degrees of
freedom are associated with continuous (coherent) quantum numbers turns out to
be unavoidable to get phase space solutions describing the electronic quantum
motion at any magnetic fields. 

As an illustration, we have considered  the motion in arbitrary quadratic 
electrostatic potentials, which is known to be exactly solvable by diagonalization 
of the Schr\"{o}\-din\-ger's equation.  The full phase space formulation (unusual in
condensed matter when dealing with fully quantum problems) offers an original
viewpoint, with a limpid underlying classical physics, on the quantization
processes, which is very different from that provided by the conventional
(historical) derivations \cite{Fock1928,Darwin1931,Fertig1987} based on the wave
function formalism. Especially, thanks to the overcompleteness of the coherent
state representation, it yields a generic (unique) solution capable to embrace
all types of quadratic potential within a simple compact mathematical
expression, without having recourse to the properties of special orthogonal
(Hermite, Laguerre, etc...) polynomials or special functions as usually 
required via the wave function formalism. We have also investigated a 
simplified model of Landau breakdown in the case of an inverted parabolic 
potential, showing a surprisingly rich phe\-no\-me\-no\-lo\-gy. Most markedly, this model displays three distinct physical stages when varying the field amplitude, in a very similar way to the situation encountered in disordered two-dimensional electronic gases.

A possible application of the full phase space formalism beyond the case of
quadratic potentials may be the derivation  of approximate functionals for the
local density of states valid in a broader magnetic field range than
originally devised in Refs. \cite{Champel2009,Champel2010} for a smooth
disordered electrostatic potential. In particular, one may expect to get
specific signatures of Landau level mixing in the characteristic features of the
effective guiding center motion. The extension of the phase space formalism may
also be useful for the study of the correlations of the local density of states
in a broader regime than in Ref. \cite{Champel2011} which neglects Landau level
mixing. However, the present phase space formulation, which naturally allows one
to perform semi-classical (local) approximations, is usually not convenient for
the study of nonlocal transport properties, which require controlled
approximations of the quantum solution on long time scales.


\appendix

\section{Dyson equation in the bicoherent state representation}

The aim of this appendix is to prove Eqs. (\ref{Green2})-(\ref{Dysonrho}).  We
first express the vortex Green's functions $g_{n_1,n_2}({\bf R};t)$ in terms of
the bicoherent Green's functions $  g_{{\bm \rho}_1,{\bm \rho}_2}\left({\bf
R};t\right)$ by inverting the relation (\ref{defgrho}) thanks to the analytical
dependence on the variables $\zeta_1^{\ast}$ and $\zeta_2$

\begin{eqnarray}
 \hspace*{-2.5cm} g_{n_1,n_2}({\bf R};t)=  \int \!\! \frac{d^2 {\bm \rho}_1}{ 2\pi l_B^2} \int \!\! \frac{d^2 {\bm \rho}_2}{ 2\pi l_B^2} \, \left(\frac{\zeta_1}{\sqrt{2} l_B}\right)^{n_1} \, \left(\frac{\zeta_2^{\ast}}{\sqrt{2} l_B}\right)^{n_2} 
\frac{g_{{\bm \rho}_1,{\bm \rho}_2}\left({\bf R};t\right)}{\sqrt{n_1! \, n_2!}}   \, e^{- \frac{|\zeta_1|^2+|\zeta_2|^2}{4 l_B^2}}. \hspace*{0.5cm}
\label{gcoherent}
\end{eqnarray}
This expression is then inserted into Eq. (\ref{Green}), which reads after
summing over the integers $n_1$ and $n_2$
\begin{eqnarray}
\hspace*{-2.5cm}
G({\bf r},{\bf r'};t)=\int \!\! \frac{d^2 {\bf R}}{2 \pi l_B^2} \int \!\! \frac{d^2 {\bm \rho}_1}{2 \pi l_B^2}  \int \!\! \frac{d^2 {\bm \rho}_2}{2 \pi l_B^2}  \,
 g_{{\bm \rho}_1,{\bm \rho}_2}\left({\bf R};t\right) 
e^{-(l_B^2/4) \Delta_{{\bf R}}} \left[ \langle {\bm \rho}_2,{\bf R}|{\bf r}' \rangle \, \langle {\bf r} |  {\bm \rho}_1,{\bf R}\rangle \right] 
 \,. \label{inter1}
\end{eqnarray}
We then reorganize the variables of integrations ${\bm \rho_1}$ and ${\bm
\rho_2}$ in the set of variables ${\bm \rho}=(\rho_x,\rho_y)$ with
$\rho_x=(\rho_{1x}+\rho_{2x})/2-i(\rho_{2y}-\rho_{1y})/2$ and
$\rho_y=(\rho_{1y}+\rho_{2y})/2+i(\rho_{2x}-\rho_{1x})/2$, and ${\bm
\rho}_{-}={\bm \rho}_2-{\bm \rho}_1$.  Introducing the change in function 
\begin{eqnarray}
 g_{{\bm \rho}_1,{\bm \rho}_2}\left({\bf R};t\right)
=\langle {\bm \rho}_1 | {\bm \rho}_2 \rangle \,  e^{i \left( {\bm \rho}_{-}  \times \hat{{\bf z}} \right) \cdot {\bm \nabla}_{{\bm \rho}} } \, e^{(l_B^2/4) \Delta_{{\bm \rho}}} g({\bm \rho},{\bf R};t) , \nonumber \\ \label{change}
\end{eqnarray}
and noting that
\begin{eqnarray}
\langle {\bm \rho}_1 | {\bm \rho}_2 \rangle \,
 \langle {\bm \rho}_2,{\bf R}|{\bf r}' \rangle \, \langle {\bf r} |  {\bm \rho}_1,{\bf R}\rangle
= e^{-\frac{{\bm \rho}_{-}^2}{2 l_B^2}} \, \langle {\bm \rho},{\bf R}|{\bf r}' \rangle \, \langle {\bf r} |  {\bm \rho},{\bf R}\rangle
, \nonumber \\
\end{eqnarray}
we then perform the integration over the variable ${\bm \rho}_{-}$  in Eq. (\ref{inter1}) to get the expression 
\begin{eqnarray}
\hspace*{-2.5cm}
G({\bf r},{\bf r'};t)=\int \!\! \frac{d^2 {\bf R}}{2 \pi l_B^2} \int \!\! \frac{d^2 {\bm \rho}}{2 \pi l_B^2}    \,
 e^{-(l_B^2/4) \Delta_{{\bm \rho}}} \left[ g\left({\bm \rho},{\bf R};t\right) \right]
\,
e^{-(l_B^2/4) \Delta_{{\bf R}}} \left[ \langle {\bm \rho},{\bf R}|{\bf r}' \rangle \, \langle {\bf r} |  {\bm \rho},{\bf R}\rangle \right] 
.   \hspace*{0.5cm}
\end{eqnarray}
Integrating by parts, we finally arrive at the result written in Eq. 
(\ref{Green2}).

The equation obeyed by the function $g({\bm \rho},{\bf R};t)$ is obtained by
projecting Dyson Eq. (\ref{Dyson}) onto the bi-coherent state representation.
After summing over the discrete Landau level indices we obtain
\begin{eqnarray}
\hspace*{-2.5cm}
\left(i \hbar \partial_t\pm i 0^{+} 
\right)
g_{{\bm \rho}_1,{\bm \rho}_2}\left({\bf R};t\right)  
-
\int \!\! \frac{d^2 {\bm \rho}_3}{ 2\pi l_B^2} \, w_{{\bm \rho}_1,{\bm \rho}_3}\left({\bf R}\right) \star_{{\bf R}} g_{{\bm \rho}_3,{\bm \rho}_2}\left({\bf R};t\right)
=\langle {\bm \rho}_1  | {\bm \rho}_2 \rangle \, \delta(t),  \hspace*{0.5cm}\label{tosolve}
\end{eqnarray}
where
\begin{eqnarray}
\hspace*{-2.5cm}
 w_{{\bm \rho}_1,{\bm \rho}_2}\left({\bf R}\right) = \hbar \omega_c \left(\frac{\zeta_1^{\ast} \zeta_2}{2 l_B^2}+ \frac{1}{2} \right)\langle {\bm \rho}_1 |{\bm \rho}_2 \rangle 
+\int \!\! d^2 {\bf r} \, 
e^{-(l_B^2/4) \Delta_{{\bf R}}} \left[ \langle {\bm \rho}_1,{\bf R}|{\bf r} \rangle \, \langle {\bf r} |  {\bm \rho}_2,{\bf R}\rangle \right] \, V({\bf r}).
\nonumber
\end{eqnarray}
The first term in the right-hand side of this latter expression corresponds to
the rewriting of the Landau level kinetic energy contribution in the bicoherent
state representation. Using the general dependence (\ref{defgrho}) of $g_{{\bm
\rho}_3,{\bm \rho}_2}({\bf R};t)$ on the variables ${\bm \rho}_2$ and ${\bm
\rho}_3$ to perform the integrals over the variable ${\bm \rho}_3$  and  setting
${\bm \rho}_1={\bm \rho}_2 \equiv {\bm \rho}$ in Eq. (\ref{tosolve}),  we
derive  in a first stage a closed equation obeyed by the diagonal component
Green's functions $g_{{\bm \rho},{\bm \rho}}({\bf R};t)$ 
\begin{eqnarray}
\hspace*{-1.5cm}
\left(i \hbar \partial_t\pm i 0^{+} 
-
 w_{{\bm \rho},{\bm \rho}}\left({\bf R}\right) \right) \star_{{\bf R}} e^{\frac{l_B^2}{2} (\overleftarrow{\partial}_{\! \! \rho_x}-i\overleftarrow{\partial}_{\! \!\rho_y} )( \overrightarrow{\partial}_{\! \! \rho_x} +i  \overrightarrow{\partial}_{\! \! \rho_y})} \, 
g_{{\bm \rho},{\bm \rho}}\left({\bf R};t\right)= \delta(t). \hspace*{0.5cm} \label{presque}
\end{eqnarray}
Considering the change in function (\ref{change}), we write down in a second
stage from Eq. (\ref{presque}) a similar equation for the function $$g({\bm \rho},{\bf
R};t)=e^{-(l_B^2/4) \Delta_{{\bm \rho}}}g_{{\bm \rho},{\bm \rho}}({\bf R};t),$$
which only differs from the previous equation  (\ref{presque}) in the structure of the
infinite-order differential operator (this step is most easily done by going
temporarily to the Fourier space following the calculations detailed in Appendix
A of Ref. \cite{Champel2009}).  The final result is provided in Eq.
(\ref{Dysonrho}), where 
\begin{eqnarray}
\hspace*{-1.5cm}
E({\bm \rho},{\bf R}) =e^{-(l_B^2/4) \Delta_{{\bm \rho}}}  w_{{\bm \rho},{\bm \rho}}\left({\bf R}\right) 
=\frac{1}{2} m^{\ast}\omega_c^2 {\bm \rho}^2 
+\int \!\!  d^2 {\bf r} \,  K({\bf r},{\bf r};{\bm \rho},{\bf R}) \, V\left({\bf r} \right). \hspace*{0.5cm}
\end{eqnarray}
The contribution arising from the potential energy $V({\bf r})$ can be further
simplified, given that
\begin{eqnarray}
\hspace*{-2cm}
\int \!\! d^2 {\bf r} \,  K({\bf r},{\bf r};{\bm \rho},{\bf R}) \, V\left({\bf r} \right) 
&=&e^{-(l_B^2/4)( \Delta_{{\bf R}}+\Delta_{{\bm \rho}})} \int \!\! \frac{d^2 {\bf r}}{2\pi l_B^2} \, e^{-\frac{\left({\bf r}-[{\bm \rho}+{\bf R}] \right)^2}{2 l_B^2}} V({\bf r}) \nonumber  \hspace*{1.68cm} \\
&=& e^{-(l_B^2/4)( \Delta_{{\bf R}}+\Delta_{{\bm \rho}})}  e^{(l_B^2/2)\Delta_{{\bm \rho}+{\bf R}}}  \,  V({\bm \rho}+{\bf R}) =  V({\bm \rho}+{\bf R}). \nonumber
\end{eqnarray}
 This means that the quantity $E({\bm \rho},{\bf R})$ is nothing but the
classical expression for the total energy.

\end{document}